\newcommand{\al}{\mbox{$\alpha $}}

\newcommand{\s}{\mbox{$\sigma $}}

\documentstyle[epsf,12pt,epsfig]{article}

\newcommand{\ls}{\mbox{$ l_{s} $}}

\newcommand{\lj}{\mbox{$ l_{p}^{11} $}}
\newcommand{\be}{\begin{equation}}
\newcommand{\br}{\begin{eqnarray}}
\newcommand{\ee}{\end{equation}}
\newcommand{\er}{\end{eqnarray}}

\newcommand{\p}{\mbox {$ \partial$}}

\begin{document}
\title{
\hfill\parbox{4cm}{\normalsize IMSC/98/05/25 \\hep-th/9805126}\\ 
\vspace{2cm}
The Hagedorn Transition and the Matrix Model for Strings}
\author{ B. Sathiapalan \\
{\em Institute of Mathematical Sciences}\\
{\em Taramani}\\
{\em Chennai 600113}\\
{\em INDIA }}
\maketitle
\begin{abstract}
We use the Matrix formalism to investigate what happens to
strings above the Hagedorn temperaure. We show that it is not a
limiting temperature but a temperature at which the continuum string picture
breaks down. We study a 
collection of $N$  D-0-branes arranged to form a string having 
$N$ units of light cone momentum.
We find that at high
temperatures the favoured phase is one where the string 
world sheet has disappeared and the low energy degrees of freedom 
consists of $N^2$ massless particles (``gluons''). The nature
of the transition is very similar to the deconfinement transition
in large-N Yang Mills theories. 
\end{abstract}
\newpage
\section{Introduction}
  
A feature of string theory that has not been fully understood is the 
existence of a ``limiting'' temperature, often called the ``Hagedorn''
temperature \cite{H,Fr,C}. The reason for a limiting 
temperature is  the exponential growth with energy
of the number of states in a relativistic string which makes 
 the canonical
partition function diverge at this temperature.  
 A lot of work was done in
exploring various aspects of this phenomenon [\cite{EA}-\cite{SO2}]. 
The physical picture that
has emerged is roughly  as follows . Consider a gas 
of weakly interacting strings. As the Hagedorn temperature is reached 
it is entropically favoured for most of the energy to go into one big
string. As a result the temperature of the string gas (measured by the
average kinetic energy of the strings) does not increase - hence the
idea of a limiting temperature.
The problem is subtle because, as noted above,
the canonical ensemble does not make
sense near the Hagedorn temperature. The partition function diverges,
the specific heat is negative and a microcanonical description has to be
used to arrive at this picture. 

What is not clear from this analysis is whether the Hagedorn
temperature
is really a limiting temperature or whether it is to be interpreted as
a phase transition temperature. One motivation for interpreting
it as a phase transition comes from the belief that large-N Yang-Mills
theory has a string description, where the string coupling
constant, $g_s$, can be identified with $1/N$.  This theory is believed to have
a transition (for any $N$) from a low temperature confining phase
where the 
lightest
degrees of freedom are colour singlets, to a high temperature
deconfined
phase where the light degrees of freedom are gluons (and quarks). 
It is only natural to identify the Hagedorn temperature with the
temperature
of the deconfinement transition.

Evidence for the existence of two distinct phases in
Yang-Mills
theory, in the large $N$
limit,
 was given in
\cite{Th}. It was shown that in the high temperature phase
asymptotic freedom requires that there be a term of O($N^2$) in the
free energy which is absent in the low temperature phase where
the degrees of freedom are colour singlets.
 It was also noted there that a possible explanation for the different
form of the expression for the free energy is the non-analyticity
caused by an 
exponential (in energy) increase in the number of states in the low
temperature phase leading to a Hagedorn temperature.

A further step in this direction was made in \cite{AW}. There is an
interpretation of the Hagedorn transition as a Kosterlitz-Thouless
transition on the string world sheet induced by a condensation of
vortices \cite{BS1,IK}. The vortices correspond to strings wound
around a compact Euclidean time coordinate. The Kosterlitz-Thouless
point is where this winding mode becomes massless.
Using this picture it was argued in \cite{AW}, that the expression for
the free energy ($- ln Z$) acquires an additional term
of O($\frac{1}{g_s^2}$) = $O(N^2)$ after the phase transition. 
Using the results of \cite{Th}
this made the connection with the deconfinement transition
stronger. Furthermore this extra term is a temperature dependent genus zero
contribution,
 which cannot ordinarily
be there if a continuum string world sheet exists. Thus the world
sheet
must somehow disappear (this can happen due to a condensation of
vortices) and this would correspond in the gauge theory to a 
 liberation of gluons. Thus a consistent physical interpretation of
the Hagedorn transition, based albeit on indirect arguments and
analogies, can be made.

Issues such as the phase above the Hagedorn
transition where the string supposedly disintegrates cannot really be 
discussed using string perturbation theory. However, in the last
year or two
a formalism that purports to treat string theory in a non-perturbative
fashion has been proposed \cite{BFSS,IKKT} and various issues
have been studied [\cite{FKKT}- \cite{muk}. 
Finite temperature
aspects of this formalism have also been discussed \cite{MOP}. 
  It is appropriate
therefore to investigate the Hagedorn transition once again in the light of
new
insights that have been obtained using this formalism. This is the
subject
of this paper. 
We will show that in the matrix model of $N$ D-0-branes, at low
temperatures, there is a string phase where the D-0-branes are spread
out to form a membrane wound around a compact direction. 
(This picture has also been used in \cite{CM} to discuss other aspects
of the Hagedorn transition.)
 In this phase the ground state
must be that of a IIA string with a specified value of string tension.
The light modes are those of a string.
 At high temperatures this string ceases to
exist. The light degrees of freedom are massless modes. These are just
fluctuations about the origin of the $N^2$ matrix elements.
This phase transition can be identified with the Hagedorn transition
for the IIA string
constructed
out of the $N$ D-0-branes. \footnote{This issue has also been addressed in the
``string-bits'' formalism, with similar conclusions \cite{CT}. The
essential idea being that the string is a polymer of ``string-bits'',
and at high temperature dissociates into its constituents.}

The matrix model calculation of free energy in the low temperature
phase is suspect because we do not know the exact zero-temperature
expression. In particular one knows that there must exist a threshold
bound state for any $N$ (this has been proved for $N=2$\cite{SS,PY})
 and this is not seen in perturbation theory.
Our calculation of the (temperature dependent part of) the effective
potential which is only a one-loop approximation is therefore not good
enough in the low temperature phase. However, even though
non-perturbative effects are expected to change the number of light degrees of
freedom, one can argue that the way it does this is by reducing the
number of flat directions and this can only reduce the number of light
degrees of freedom. Thus hopefully we have overestimated the magnitude
of the free energy and therefore the existence of the phase transition
can still be argued.

 In matrix
theory a membrane is described in the limit of $N$ being large
by the following configuration \cite{BFSS}:
\be \label{memb}
X^i= L^i p ; X^j = L^j q
\ee
Here $p,q$ are matrices that satisfy \footnote{Explicit
representations can be easily found in the literature. e.g.
\cite{zach,BS}}

\be
[p,q] = \frac{2\pi i}{N}
\ee
One can consider a representation (as in \cite{BS}) where the
eigenvalues range from 0 to 2$\pi$ in $N$ steps.
(\ref{memb}) represents a membrane in the $ij$ plane, of sides
$2\pi L^i$ and $2\pi L^j$. If $i,j$ happen to be compact
dimensions of radius $L^{i,j}$, then (\ref{memb}) describes a wrapped
membrane, which classically describes a stable BPS configuration. 
We can also consider a background configuration of the form \cite{BS2}
\be     \label{str}
X^i = L^i p
\ee
which describes a membrane wrapped around $X^i$ with the other edge
free i.e. a free string, not necessarily wrapped. If we let
the $X^j$ be periodic functions of $q$ of the form $e^{imq}$, then
we get a closed string. Let us construct this string action in the
matrix model.

    The matrix model for M-theory \cite{BFSS} is essentially a
D-0-brane 
action,
reinterpreted as a DLCQ M-theory \cite{Suss}  where the ``eleventh''
dimension, of radius $R^{-}$, is a light cone direction. This is
related \cite{NS} to $R_{11}$ of the M-theory that we would like to
study
by a very large boost.The
$P^{+}$ component of momentum is 
 $\frac{N}{R^{-}}$ (where $N$ is the number of D-0-branes),
 and is like a ten dimensional mass. When
$P^{+}$
is large (as it is in an infinite momentum frame) the action is that
of a non relativistic particle with mass $\frac{N}{R^{-}}$.
Thus the evolution operator  $P^{-} = \frac{P_{tr}^{2} +m^{2}}{2P^{+}}$
is the Hamiltonian obtained from the D-0-brane action.
The bosonic part of the action for N D-0-branes
\footnote{ This supersymmetric action has also been studied
in a different context by \cite{halp}}
is
\be  \label{2.1}
S=
\frac{1}{2g_{s}}\int \frac{dt}{\ls}Tr\{(\p _{t} X^{i})^{2} +
\frac{1}{4\pi ^{2}\ls
^{4}}[X^{i},X^{j}]^{2}\}
\ee

We have explicitly written factors of $\ls$, to make the action
dimensionless.  Our conventions are as follows: $g_{s}$ is the
(IIA) string coupling constant and
we have defined the inverse string tension to be $2\pi \al ' $
with $\al ' = \ls ^{2}$.  The parameters of DLCQ M-theory are the
radius $R^{-}$ and the eleven dimensional Planck length $\lj $. $\lj$
is defined by the membrane tension which we have taken to be
$\frac{1}{(2\pi )^{2} (\lj ) ^{3}}$.  This fixes $g_{s}^{2} =
(\frac{R^{-}}{\lj })^{3}$ and also $\al ' = \frac{(\lj )^{3}}{R^{-}}$.
These relations also imply that $g_{s} \ls = R^-$. Thus the kinetic
term is essentially ``$\frac{1}{2} mv^{2}$'' with $m =\frac{1}{R^{-}}$.
The coefficient of the potential term can be seen to be $\frac{R^{-}}
{8\pi ^{2} (\lj )^{6}}$.
The normalization of the potential term is chosen so that the 
classical mass of
a D-2 brane comes out right. The parameters of M-theory are related to
those of this DLCQ matrix theory by the scaling relations \cite{AS,NS}
$\frac{R^-}{(\lj )^2} = \frac{R_{11}}{(\tilde{\lj})^2}$,
$\frac{L^9}{\lj} = \frac{\tilde{L^9}}{\tilde{\lj}}$.

Let us assume that $X^9$ is a compact dimension of radius $L^9$,
assumed to be small. We will take this as the ``eleventh dimension''.
 When a membrane is wrapped around $X^9$ we
get a string with inverse tension $\beta '$ = $\frac{(\lj )^3}{L^9}$,
and string coupling $g_{s\beta '}^2$ = $\frac{(L^9)^3}{(\lj )^3}$.

The 1+1 dimensional action that we obtain is 
  that of  9+1 dimensional Supersymmetric $U(N)$
Yang-Mills theory reduced to 1+1 dimension.
\[      
S=
\frac{1}{2g_{s}}\int \frac{dt}{l_{s}}\int _{0}^{2\pi L_{9}^{*}}
 \frac{dx}{2\pi L_{9}^{*}}
Tr\{ (\p _{t}X^{i})^{2} - (D_{x}X^{i})^{2} + 
\]
\be     \label{2.1.1}
+ (F_{09})^{2}
+\frac{1}{4\pi ^{2}l_{s}^{4}}([X^{i},X^{j}])^{2}\}
\ee

$D_{x} =  \p _{x} +i A_{9}$ is the covariant derivative in a 
direction $X^{9*}$, which is T-dual to $X^{9}$, and is of radius
$L_{9}^{*}$. $x$
is thus a coordinate along a D-1-brane wound around $X^{9*}$.
If $X^{i}_{mn}$ is a matrix describing the lattice of D-0-branes, ($m,n
=0...M$) then
following \cite{WT}, we have
\be     \label{2.2.2}
A^{9} = \frac{1}{2 \pi \al '} \sum _{n=0}^{M} e^{inx\frac{L_{9}}{\alpha '}}
X_{0n}^{9}
\ee 
Note that $X_{00}$ is the original  D-0-brane matrix of the
uncompactified
theory. For us $X^9 _{00}$ will be given by (\ref{str}). Thus $D_x$ is
given by

\be
\p _x \otimes I + I \otimes \frac{L^9}{\alpha ' N}\p _q
\ee
The derivatives act on eigenfunctions:
\be    \label{ef}
e^{ir\frac{x}{L_{9}^{*}}}e^{imp}e^{inq}
\ee
 The eigenvalues are $ \frac{rN + n}{NL_{9}^{*}}$. Thus the effective
radius is $NL_9^*$. These are the long strings described in 
\cite{DM,MS,Motl,Ver}.
 We can replace $D_x$ by $\p _{\sigma}$ and the eigenfunctions (\ref{ef})
by $e^{\frac{iu\sigma}{NL^{9*}}}$. Here $\sigma$ has a range of $0-2\pi
NL^{9*}$. 
 Using $L_9^* = \frac{\alpha '}{L^9} = \frac{\beta
'}{R^-}$
we can rewrite the action (\ref{2.1.1}):
\be \label{lc}
\frac{1}{4\pi  \beta '}\int dt \int _{0}^{\frac{2\pi N \beta '}{R^-}}d\s
\{ (\p _t X^i)^2 - (\p _ {\sigma}X^i )^2 +...\}
\ee
Noting that $\frac{N}{R^-}$ is $P^+$, we recognize the light cone
gauge string action with a string tension of $\beta '$.
Turning on $F_{09}$ corresponds to addition of D-strings \cite{EW}.
The commutator terms are zero if we restrict the matrices $X^i$ to be
$X^i(x,q,t)$ i.e. without any $p$ dependence.  $p$ dependence corresponds to
fluctuations in the matrix model that are not string-like. If we knew 
the fully non-perturbative effective action presumably it would be
manifest that these fluctuations are massive.

If one calculates the partition function using the above string action
one will encounter the usual divergence as the Hagedorn temperature
is reached.  However we see that the divergence is an artifact. At the
Hagedorn temperature energy in the string is overwhelmingly found 
in the highest
mode that can be excited with that energy. In the matrix model there
is a cutoff on the mode number which is in fact $NM$. Thus in the
partition function there is a natural cutoff in the energy integral
above which the discrete nature of the string becomes important and
at this point one has to go back to the original matrix model. Thus
we see that the Hagedorn temperature is not a limiting temperature.
Of course eventually one has to let $NM \rightarrow \infty$ (in fact one should
sum over all values of $N$).  Presumably some sort of scaling has to
be done to extract finite physical quantities, as has been discussed
in, for example, \cite{IKKT,FKKT,BS,BS2}. We should also keep in mind that 
$\tilde{\beta} ' = \beta ' (\frac{\tilde {\lj}}{\lj})^2$.  Thus 
$\beta ' \rightarrow \infty $ as $\tilde {\lj}\rightarrow 0$, in order
to get a finite $\tilde {\beta}'$.

Let us try to understand the nature of the phase transition.
We have shown above the emergence of a string in the matrix model
when the eigenvalues of a particular $X^i$ are spread over a finite 
range. $L$ can thus be thought of as an order parameter characterizing
the two phases of the model.
  When $L$ is non-zero
we have a configuration where the eigenvalues of $X^i$ are uniformly
distributed from 0 to $2\pi L$ and it describes $N$ D-0-branes
arranged to give a string, whereas if $L$ is zero we get a cluster
of $N$ D-0-branes located at the origin and no string. Given this,
there is a simple way to understand the nature of the phase transition.
 The issue of
whether a uniform distribution of eigenvalues is the preferred state
or not goes back to the issue of the breaking of ``$U(1)^d$''  symmetry
in the reduced Eguchi-Kawai models [\cite{EK}-\cite{GK}]. A phase transition of
this type as a function of the coupling constant in  large-N lattice
gauge
theory was also noted in \cite{GW,SW}.
In non supersymmetric
models even in the one loop approximation one can see that there is
a logarithmic attractive potential between the eigenvalues and they
tend to cluster, thus breaking the $U(1)^d$ symmetry. 
In supersymmetric models on the other hand, Bose-Fermi
cancellation allows a uniform distribution \cite{IKKT}. Thus if one 
assumes that supersymmetry somehow prevents the clustering of
eigenvalues, then one is led immediately to conclude that at finite
temperature, supersymmetry being broken, the eigenvalues will start
clustering. This then is the basic mechanism of the Hagedorn
transition.

Of course the above argument does not explain why there should be 
a {\em finite} transition temperature - supersymmetry is broken
for arbitrarily low temperature. This clearly has to do with the
uniqueness of the assumed ground state. Somehow the flat directions
should be lifted and a non-trivial potential generated. This will
ensure that only at a finite temperature will the temperature effects
dominate and change the nature of the equilibrium (themodynamic) state.
Thus one needs to know the exact potential to make this argument
concrete.

We do not know how to compute the exact potential but it is sufficient
for our purposes that the configuration (\ref{str}) should have non-zero
overlap with the (unique)
ground state and therefore lies in the ``bound '' region of whatever
non-trivial potential for $L$ the exact theory has.  The argument
for this is as follows:
  We know that for $N$ D-0-branes there is exactly one
threshold bound state. We also know that this is the (super)graviton
with $N$ units of momentum along $R^{11}$. From the viewpoint of the
10 dimensional theory where $X^9$ is the eleventh direction and $X^{11}$
is the light-cone direction, this is the graviton with $N$-units
of light-cone ($p^{11}$) momentum. But this graviton is precisely the
ground state of the light-cone string described by (\ref{str}).
Therefore this configuration must have a non-zero overlap with the
 unique non-degenerate ground state. Thus if $L$ denotes the spread
of eigenvalues, which is the size of the bound state, the ground state
wavefunction denoted by $\psi (L)$ is non zero for a range of values
of $L$ and therefore $<L^9 | \psi > \neq 0$.  Here
the ket $|L^i >$ denotes the state corresponding to the classical
configuration given by (\ref {str}).
The above considerations thus explain why the Hagedorn transition
happens at a finite temperature.

We now turn to a computation of the temperature dependent free energy.
It will be a one loop computation about two different backgrounds, one
where $L$ is non-zero and one where $L=0$. We will assume that one
dimension is compact, say $X^9$ , with radius $L^9$.  
We will use the standard trick of compactifying the time coordinate
$t$ in the matrix model to a range of $\beta $ (not to be confused
with the inverse tension $\beta '$). Of course $t$ is conjugate to
$P^-$ rather than $H$, but in the infinite momentum limit for the
10-dimensional theory it is the non-relativistic energy and therefore $t$ is a
non-relativistic approximation to the time coordinate. The calculation
of $ln Z$ is straightforward (\cite{IKKT}. We have to keep in mind
that 
fermions
have anti-periodic boundary conditions in time.
The bosonic contribution is,
\be
5 Tr \;ln \;E
\ee
The fermionic contribution is 
\be
-4Tr \;ln \;E 
\ee
and the ghost contribution is 
\be
-1 Tr \; ln \;E
\ee
  where $E = P_0^2 + P_9 ^2 + b^2$.  Here $P_0 =\frac{m}{\beta}$
for bosons and ghosts and $P_0 = \frac{m+1/2}{\beta}$ for fermions.
$P_9 = \frac{n}{R}$ for all the fields.   $R$ is some length
(which could be $L^{*9}$ or $NL^{9*}$ for instance). $b$ is some
small parameter needed to regularize divergences
and which will normally be set to zero.          
Thus we obtain
\be
4N\sum _{n,m} \int _0 ^{\infty} \frac{ds}{s}(
e^{-({m^2 \over \beta ^2}+ {n^2 \over R^2})s} \; - \;
e^{-({(m+1/2)^2 \over \beta ^2}+ {n^2 \over R^2})s})
\ee
The factor $N$ comes from the degeneracy due to the factor
$e^{irp}$ (where $r$ has a range from $0$ to $N-1$) in the eigenfunctions 
(\ref{ef}).\footnote{As mentioned earlier the factor N is an
overestimate. The exact potential would break the remaining degeneracy
and should make it O(1).}
 
 Using a Poisson resummation on $m$ one reduces this to
\[
-4N{\sqrt \pi} \beta \sum _n \sum _{p \; odd }\int  _{0}^{\infty} 
\frac{ds}{s^{3\over 2}}e^{-{n^2 \over R^2}s} e^{-\frac{\pi ^2 \beta ^2 p^2}{s}}
\]
\be
= \; -8N \sum _{p \; odd \; ,n} {1 \over p}e^{-2\pi {\beta \over R} pn}
\ee
When $\beta \rightarrow \infty$, terms with $n\neq 0$ clearly go to
zero. When $n=0$ one has to introduce the regulator $b$ (which replaces
$n \over R$ in the above formula) and take the limit $b \rightarrow 0$
after taking the zero temperature limit and again the expression vanishes.
This is as it should be indicating that at zero temperature Bose-Fermi
cancellation makes the one loop correction vanish.

The zero mode term $n=0$ gives a logarithmic divergence due to the sum
over $p$.  On regularizing with $b$ we get $8Nln \; b\beta $.  For large
$\beta \over R$ the rest of the contribution is $-8Ne^{-\frac{2\pi\beta}{R}}$.
$R$ in the low temperature case is $NL^{9*}$.
Thus we get 
\be
\beta F = N(8ln \; \beta b - 8e^{-2\pi \beta \over NL^{9*}}) + \;
higher \; orders 
\ee
The logarithmic term can be seen to be the classical (high
 temperature)
 term in the free
energy of a harmonic oscillator  (any temperature is high
 relative to zero!).\footnote{The zero mode 
corresponds to a free
particle  whose partition function (in one dimension) goes as $\sqrt
{Tm}V$ where $V$ is the (infinite) volume . 
However when we  add the $b$-term (a harmonic oscillator
potential term), it regulates this infrared divergence and contributes
a factor 
of $\sqrt {T\over k}$ (in place of $V$) where
$k$ is the ``spring constant '', and the net
contribution is $T\over b$. } The second term is just the leading term
in a low temperature expansion of the free energy of a bunch of
harmonic oscillators which goes to zero
as $\beta \rightarrow 0$. Thus at low temperatures it behaves like a
zero-dimensional point particle rather than a one-dimensional field theory.

At high temperatures (small $\beta$) we get
\be	\label{ht}
\beta F = N(8ln \; (\beta b) + 8ln \; {\beta \over R} -\frac{\pi RT}{2}) +... 
\ee
The leading behaviour is thus that of a classical one-dimensional
field theory.

We can consider two extreme cases: $R= NL^{9*}$, which corresponds
to a string configuration with the order parameter $L$ being equal to
$L^{9*}$.  In that case (\ref{ht}) becomes 
\be	\label{O}
\beta F = N(2ln \; (\beta b) + ln \; {\beta \over NL^{9*}} -
\frac{\pi NL^{9*}T}{2})  +...
\ee

We can see the connection with string thermodynamics as follows:
From the above result for the free energy,
 \[
<P^->= \frac{E^2}{P^+} \approx NL^{9*}T^2 
\] 
and
this gives $E \approx {\sqrt \frac{N^2 L^{9*}}{R^-}} T$. Using
$S \approx NL^{9*}T$ (and $L^{9*}R^- = \beta '$) we get
$S \approx {\sqrt \beta '} E$, characteristic of strings.

In the case where all the D-0-branes are clustered at the origin and 
$L=0$, we have to set $R= L^{9*}$.  
There is one other difference with the previous
case.  The degeneracy
of each level is $N^2$ rather than $N$.  This is because the
eigenvalues of $D_x$ acting on the eigenfunctions given in (\ref{ef})
do not depend on $n,m$  (i.e. the eigenfunctions can have any $p,q$
dependence). Thus the entire expression for the free energy is now
multiplied by $N^2$. This gives
\be	\label{DO}
\beta F = N^2(2ln \; (\beta b) + ln \; {\beta \over L^{9*}} -
\frac{\pi L^{9*}T}{2}) +... 
\ee
Notice that the leading high temperature piece is the same in both cases.
Both are proportional to $N^2$. However the non-leading term is different. 
Clearly for small $\beta$ when the logarithm becomes negative 
(\ref{DO}) is more negative than (\ref{O}).
 Hence the phase without the string is
favoured at high temperatures. 

Let us summarize what we have learnt. The phase below the Hagedorn
temperature corresponds to one where the D-0-branes are arranged to
form a membrane which can be wrapped around a compact direction to
give
a string.  At high temperatures the D-0-branes prefer to cluster at
one point, thus the string disappears.  We cannot estimate the
temperature at which this happens without some knowledge of the
exact potential as a function of $L$ \footnote{An attempt
to calculate the potential is described in \cite{DFS}}.
 If the height of the potential
at $L=0$ is of order $N^2 \over \sqrt {\beta '}$, then the transition
can be expected to take place at around a temperature of
$\frac{1}{\sqrt \beta '}$ which is essentially the Hagedorn
temperature. But there is no estimate of this in the literature as
far as we know.

The other thing we have learnt is that the nature of the phase
transition
is very similar to the deconfinement transition in large-N Yang-Mills.
In \cite{pol} it was shown that the deconfinement transition has
exactly the same form. The two phases described there were the same:
One where the eigenvalues are uniformly distributed and one where they
are clustered at one value. Furthermore the fact that at high energies
the free energy has a term of O($N^2$) and that it corresponds to
massless particles in the adjoint representation (``gluons'') also
makes the connection more concrete.

The spacetime picture above this phase transition is still a little
mysterious. Does the cluster of D-0-branes at the origin still represent
gravitons or perhaps a black hole? That is reminiscent of situations
discussed in \cite{HP}. Also, the membrane of M-theory can wind around
any compact
dimension and give a stable string configuration. In particular it can
wind around $X^{11}$.  These strings are the ones that connect the
D-0-branes. As the matrix model does not contain these string
excitations, (indeed the scaling prescription of \cite{AS,NS} ensures
this), we cannot discuss them. 
Nevertheless if 
we interpret the results of this paper to mean that the 
membrane configuration itself is unstable then there cannot be any 
strings left at all. 
It is also possible that there are other
phase transitions at temperatures $T \approx O( 1/R_{11})$,
which cannot be computed in this model.

\noindent
{\bf Acknowledgements}\\
I would like to thank N. D. Hari Dass and S. Kalyana Rama for several
extremely useful discussions.

\end{document}